\documentclass[notitlepage]{revtex4-1}

\usepackage[utf8]{inputenc}
\usepackage{graphicx}
\usepackage{amsmath}
\usepackage{amsfonts}
\usepackage{xcolor}

\let\vec\mathbf

\begin{document}

\title{Static properties of quasi-confined hard-sphere fluids}

\author{Charlotte F. Petersen}
\email[]{charlotte.petersen@uibk.ac.at}
\author{Lukas Schrack}
\author{Thomas Franosch}\address{Institut f{\"u}r Theoretische Physik, Leopold-Franzens-Universit{\"a}t Innsbruck, Technikerstra{\ss}e 21A, A-6020 Innsbruck, Austria.}

\begin{abstract}
Confined fluids display complex behavior due to layering and local packing. Here, we disentangle these effects by confining a hard-sphere fluid to the surface of a cylinder, such the circumference extends only over a few particle diameters. We compare the static structure factor and the pressure measured in computer simulations to the Percus-Yevick closure in liquid state theory. A non-monotonic evolution of the static-structure-factor peak and the pressure is observed upon variation of the confining length, similar to a liquid confined between two plates. This indicates that the density profile and the particle correlations may not be intrinsically connected in real confined liquids.   
\end{abstract}

\maketitle


\section{Introduction}\label{sec:intro}

Confined liquids are prevalent in nature~\cite{beebe2002physics} and have many industrial applications~\cite{urbakh2004nonlinear,pinilla2005structure}. Physical confinement is known to affect fundamental properties of a fluid, such as diffusion~\cite{mittal2008layering}, the glass transition~\cite{varnik2016non,lang2010glass,krakoviack2005liquid,krakoviack2005liquid2}, and the freezing transition~\cite{schmidt1996freezing,schmidt1997phase}. Confinement induces ordering of the particles, which results in density modulation~\cite{hansen2013theory,bunk2007confinement}. However, the local properties of the confined fluid are not simply related to bulk fluids with equivalent densities~\cite{goel2008tuning,mittal2008layering,goel2009available,bollinger2015communication}. A slit geometry, where a fluid is confined between two parallel walls, has been extensively used in investigations into the effects of confinement experimentally~\cite{nugent2007colloidal,nygaard2012anisotropic,nygaard2013local,nygaard2016anisotropic,nygaard2017anisotropic,C7CP02497E}, theoretically~\cite{deb2011hard,mandal2014multiple, C7SM00905D}, and also computationally~\cite{mittal2008layering,PhysRevLett.118.065901,geigenfeind2015confinement}.
In this layered fluid, the density oscillates as a function of distance from the walls. 
For extreme confinement, diffusion is slowed down dramatically compared to the bulk~\cite{scheidler2004relaxation,gokhale2016localized,eral2009influence} and  remarkably, both the diffusivity and the glass transition show a non-monotonic dependence on the plate-to-plate distance~\cite{mandal2014multiple,lang2010glass,lang2012mode,lang2013mode,lang2014glassy}. These non-monotonic effects are a result of a subtle interplay between the structure of the local cage created by neighbors, and the layering induced by the walls. There exists empirical evidence that these unusual long-range transport properties are intrinsically related to structure, through a scaling between diffusivities and excess entropies~\cite{ingebrigtsen2013predicting}. Similarly,  the mode-coupling theory of the glass transition~\cite{gotze2008complex}  predicts dynamic properties using only structural information as input. Thus, the relevance of static correlations to transport properties provides a strong motivation to further investigate the structure of fluids under confinement.

Apart from the inhomogeneous density profile, the structure of a confined liquid is also characterized by its pair-distribution function or equivalently its structure factor, which also show non-monotonic behavior with confinement length~\cite{mandal2014multiple}. The structure factor is a key measure of the correlations in a fluid, as it can be calculated in computer simulation, theoretically, and measured directly in scattering experiments. In liquid state theory a common approach is to use integral equations to formulate closure relations, the simplest being  
the Percus-Yevick (PY) approximation~\cite{hansen2013theory,percus1958analysis,thiele1963equation,wertheim1963exact}. In bulk hard-sphere fluids an analytic solution of the structure factor exists in odd dimensions~\cite{freasier1981remark,leutheusser1984exact}, and aside from 3D space, it has also been studied extensively in 1D~\cite{wertheim1964analytic}, 5D~\cite{freasier1981remark,leutheusser1984exact,gonzalez1990thermodynamics}, and 7D~\cite{robles2004equation}.
In 2D the PY equations have been solved numerically~\cite{lado1968equation,leutheusser1986percus}, and more recently using a convenient method which allows the results for all densities to be calculated at once~\cite{adda2008solution}.
Unfortunately, the structure factors calculated with the PY approximation deviate somewhat from those calculated accurately with computer simulations~\cite{hansen2013theory}. Yet, it still proves useful, since parameters can be readily changed without running new simulations and the solutions can be evaluated to arbitrary precision. 
The PY theory has been extended to confined systems, which are more complicated because they are inhomogeneous. The statics have been solved for hard spheres confined to a slit~\cite{sokol1980solution,kjellander1991pair,boctan2009hard,nygaard2013local}, compared with experiment~\cite{nygaard2012anisotropic}, and used to predict the dynamics~\cite{mandal2014multiple,lang2010glass}. 

In this confined system, the density profile changes as the degree of confinement is varied, and the correlations in the liquid display a non-trivial dependence on this confinement length. Thus a natural question arises; are the unusual correlations in the confined liquid are a mere reflection of the layering, or are they an independent manifestation of the confinement itself?  

\begin{figure*}
\includegraphics[trim={0 4cm 0 0},clip]{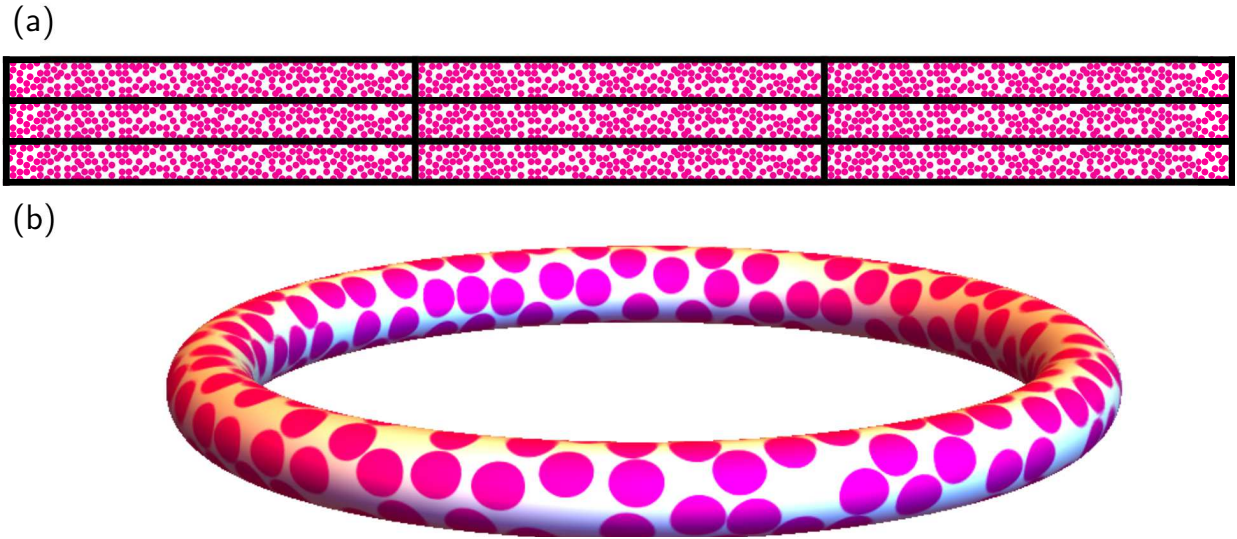}
\caption{\label{fig:torus} Illustration of the 2D analog of our simulated system. (a) Periodic simulation cells. Space is tiled periodically with one simulation cell, such that particles at the top of the cell interact with particles at the bottom of the cell, and likewise for the left and right edges. The length of one side of the periodic box (the $z$-direction) is much shorter than the other, almost bulk direction. The 3D simulations have 1 short dimension and 2 long dimensions, effectively a periodic slab. (b) The periodic cells in (a) can be represented as a single simulation cell, wrapped around the surface of a torus.}
\end{figure*}

In this paper we investigate confinement in a liquid with uniform density. We do this by considering a 3D periodic liquid, that is, one which is confined to the surface of a torus  or long cylinder. This quasi-confined fluid is illustrated in figure~\ref{fig:torus} for the two-dimensional analog. This model has the advantage of being translationally invariant, but confinement on the order of the particle diameter can still be imposed, allowing us to investigate strong confinement effects without the complications of walls and density modulation.

Similar systems have previously been used as models of confinement, for example in the study of non-equilibrium dynamics in a narrow channel~\cite{benichou2013geometry,benichou2016nonlinear}. In the turbulence community, a similar method of reducing the complexity of a system is used, referred to as Fourier-mode reduction or wavenumber decimation, to investigate the effects of local versus non-local dynamics~\cite{grossmann1996developed,meneguzzi1996sparse,de2007shear,laval2001nonlocality}. A similar periodic geometry is used in finite size scaling analysis to probe the length scales relevant to the glass transition~\cite{karmakar2009growing,PhysRevE.86.031502}, but with a cubic periodic cell, where the edge length is much larger than a particle diameter. An anisotropic unit cell geometry has been used to calculate the critical Casimir force in a binary mixture close to criticality~\cite{puosi2016direct}. In molecular dynamics simulations, finite size effects from the limited length of periodic simulation cells are known to cause anisotropy in calculated pair-distribution functions~\cite{denton1997implicit,pratt1981effects,pratt1981effects2} and stress~\cite{gonzalez2005stress}. These effects are particularly significant in non-equilibrium simulations where the periodic boundaries move~\cite{bernardi2015system,petravic1998nonlinear}. Until now, these effects have usually been considered as an undesirably consequence of finite computing resources, rather than as a deliberate method of confinement. 
A more thorough understanding of the properties of quasi-confined liquids, as the degree of confinement approaches the particle diameter, will be important to gaining a more thorough understanding of the properties of the more complex case of confined liquids. 
 
We study the static properties of quasi-confined hard-sphere fluids using theory and event-driven simulations. The quantities relevant to the study of quasi-confined fluids are presented in detail in section~\ref{sec:methods}, along with the integral equation theory and simulation methods. The results are presented in section~\ref{sec:results}, were we find a good quantitative match  between the PY theory and the simulations at low packing fractions, and semi-quantitative agreement for higher densities. For both methods we observe non-monotonic behavior of the static structure and the pressure with increasing confinement length, similar to the behavior of confined liquids in a slit geometry. In section~\ref{sec:conclusions} we discuss the conclusions of the work.
 
\section{Methods}\label{sec:methods}

\subsection{Theory of  quasi-confined fluids} 

We consider  $N$ hard spheres of diameter $\sigma$ in three dimensions   using periodic boundary conditions, where one dimension is much smaller than the other two, shown schematically in figure~\ref{fig:torus}(a). This can be thought of as spheres confined to the surface of a four dimensional torus, illustrated in 2D in figure~\ref{fig:torus}(b).
The small periodic dimension is designated as the $z$-direction, such that  $-L/2 \leq z \leq L/2$. 
We anticipate the thermodynamic limit,   $N\to \infty$, 
with lateral box size $L_{\text{box}} \to \infty$ such that the area density $n_0 = N/L_{\text{box}}^2$ remains fixed.  Thus, these fluids are a special case of the more general class of layered fluids, such as fluids next to a wall~\cite{roth2010fundamental}, in a slit geometry 
or fluids subject to a one-dimensional potential~\cite{bollinger2014structure,saw2016role,herrera2007structure,evers2013colloids,capellmann2018dense}, 
which display translational symmetry along the $x$--$y$ direction  as well as rotational symmetry around the $z$-axis.  

The structural properties of layered  fluids are encoded in averages of the multi-particle densities. Here we focus on the lowest order quantities, the simplest being the fluctuating density defined by
\begin{equation}
    \rho(\vec{r},z) = \sum_{i=1}^N \delta(\vec{r}-\vec{r}_i ) \delta(z-z_i), 
\end{equation}
where  $(\vec{r}_i,z_i)$ are the lateral and transverse  coordinates of particle $i$. For these  layered fluids the equilibrium density
$n(z) := \langle \rho(\vec{r},z) \rangle$ depends only on the transverse direction, where
$\langle \cdot \rangle$ denotes ensemble averaging.
For  our quasi-confined liquid, this equilibrium density  is even uniform
$ n(z) = n  \equiv n_0/L$.

The next simplest property is the two-particle density, 
\begin{align}
\rho^{(2)}(\vec{r},\vec{r\,}^{\prime},
z,z') &= \mathop{\sum\sum}_{i\neq j}^N
  \delta(\vec{r}-\vec{r}_i ) \delta(z-z_i)
    \delta(\vec{r\,}^{\prime}-\vec{r}_j ) \delta(z'-z_j) \nonumber \\
    &= \rho(\vec{r}, z) \rho(\vec{r\,}^{\prime},z') - \delta(\vec{r}-\vec{r\,}^{\prime}) \delta(z-z') \rho(\vec{r},z).
\end{align}
Its average for layered fluids depends only on the relative lateral distance $|\vec{r}-\vec{r\,}'|$, and this property in inherited by all two-point correlation functions. 
Then, the general definition of the pair-distribution function~\cite{hansen2013theory} reduces for layered fluids to  
\begin{align}\label{eq:pair_correlation}
g(|\vec{r}-\vec{r\,}'|,z, z') :=  \frac{\langle 
\rho^{(2)}(\vec{r},\vec{r\,}',z,z') \rangle}{n(z) n(z')} ,
\end{align} 
and 
provides readily interpretable  information on the local packing.

In contrast, 
 the density-density correlation function
\begin{align}\label{eq:density-density}
    G(|\vec{r}-\vec{r\,}'|, z, z') :=
\frac{1}{n_0} \langle \delta \rho(\vec{r},z) \delta \rho(\vec{r\,}',z') \rangle ,
\end{align}
where $\delta \rho(\vec{r},z) = \rho(\vec{r},z) - n(z)$, is the fluctuating density,  
can be measured in scattering experiments and reflects layering as well as packing. Although both quantities can be related, for general  layered fluids the density profile $n(z)$ explicitly enters. 
However, for quasi-confinement  
both quantities are trivially related, which can be made explicit in terms of the total correlation function $h:=g-1$ (which decays to zero for large distances). Then equations~\eqref{eq:pair_correlation} and \eqref{eq:density-density} yield
\begin{equation} \label{eq:total_correlation}
h(|\vec{r}-\vec{r\,}'|,z,z') =  
\frac{L^2}{n_0} G(|\vec{r}-\vec{r\,}'|,z,z') - \frac{1}{n} \delta(\vec{r}-\vec{r}') \delta(z-z') ,
\end{equation} 
and the dependence on the transverse coordinates reduces to  the relative distance $z-z'$. Furthermore, reflection symmetry reveals that it depends only on $|z-z'|$. 
Hence, the geometry of quasi-confinement, where confinement is introduced without walls or layering, is ideally suited to disentangle local packing and layering and greatly simplifies the analysis.

Here, we follow the strategy developed for the slit geometry~\cite{lang2010glass,lang2012mode,lang2013mode,doi:10.1063/1.4867284} and expand the dependence on the transverse positions in all quantities in terms of  discrete Fourier modes $\exp{\left(-i Q_\mu z\right)}$
in the  $z$-direction ($Q_\mu=2\pi\mu/L, \mu\in\mathbb{Z}$). In contrast, 
the spatial dependence parallel to the $x$--$y$ plane is decomposed into ordinary plane waves $\exp(-i \vec{q}\cdot\vec{r})$
with $\vec{q}=(q_x,q_y)$. These wave vectors are taken as discrete initially $(q_x, q_y) \in (2\pi/L_{\text{box}}) \mathbb{Z}^2$, however in the thermodynamic limit they become continuous variables, such that sums are replaced by integrals $(1/A) \sum_{\vec{q}} \ldots \mapsto (2\pi)^{-2} \int \mathrm{d}^2 q \ldots$ as usual, with $A= L_{\text{box}}^2$.  
Then the following  orthogonality and completeness relations hold
\begin{align}
 &\frac{1}{A}\int_A e^{i(\vec{q}-\vec{q\,}^\prime)\cdot\vec{r}} \mathrm{d}^2 r =
  \delta_{\vec{q},\vec{q\,}^\prime}, \\
 &\frac{1}{A}\sum_{\vec{q}}e^{i \vec{q}\cdot(\vec{r}-\vec{r\,}^\prime)} = 
  \delta(\vec{r}-\vec{r\,}^\prime), \\
 &\frac{1}{L}\int_{-L/2}^{L/2}\exp{\left[i(Q_\mu-Q_{\mu^\prime})z\right]}\mathrm{d}z = 
  \delta_{\mu,\mu^\prime}, \\
 &\frac{1}{L}\sum_\mu\exp{\left[i Q_\mu (z-z^\prime)\right]} = 
  \delta(z-z^\prime).
\end{align}

The microscopic density is now
\begin{equation}
 \rho(\vec{r},z)=\frac{1}{A}\sum_{\vec{q}}\frac{1}{L}\sum_\mu \rho_\mu(\vec{q})
  \exp{\left(-i Q_\mu z\right)}e^{-i\vec{q}\cdot\vec{r}},
\end{equation}
where the expansion coefficients
\begin{equation}
 \rho_\mu(\vec{q}) = \sum_{i=1}^N \exp{\left(i Q_\mu z_i\right)}
  e^{i \vec{q}\cdot\vec{r}_i}  
\end{equation}
take the role of the 
fundamental observables.  By translational invariance in the unbounded directions  
\begin{equation}
 \langle \rho_\mu(\vec{q}) \rangle = A n_\mu \delta_{\vec{q},0} \, ,
\end{equation}
where $n_\mu = \int_{-L/2}^{L/2} n(z) \exp( i Q_\mu z) \mathrm{d} z $ is the Fourier coefficient of the density profile. For quasi-confinement only $n_0$ is non-vanishing, corresponding to the homogeneous density. 

Translational symmetry in the $x$--$y$ direction  shows that the modes $\rho_\mu(\vec{q})$ are orthogonal for different wavenumbers. Thus the 
 correlations between such modes are characterized by  the  generalized structure factor 
 \begin{equation}
 S_{\mu\nu}(q) = \frac{1}{N}\langle \rho_\mu(\vec{q})^* \rho_\nu(\vec{q}) \rangle. 
\end{equation}
By rotational invariance around the $z$-axis the structure factors also depend only on the modulus $q= |\vec{q}|$ of the wave vector. For quasi-confinement, translational invariance along the z-direction implies that only the diagonal elements  survive  
\begin{equation}
S_\mu(q) \equiv S_{\mu\mu}(q) = \frac{1}{N}\langle |\rho_\mu(\vec{q})|^2\rangle. \label{eq:SF_sim}
\end{equation}
Furthermore, in this case $S_{-\mu}(q) = S_\mu(q)$ as a consequence of reflection symmetry at the $x$--$y$ plane. 

 The mode expansion of the density-density correlation function for layered fluids reads 
\begin{align}
G(|\vec{r}-\vec{r\,}'|,z,z') 
&= \frac{1}{A}\sum_{\vec{q}}\frac{1}{L^2}\sum_{\mu\nu}
  S_{\mu\nu}(q) e^{i \vec{q}\cdot(\vec{r}-\vec{r\,}') } 
  \exp{\left[i ( Q_\mu z - Q_\nu z') \right]},
\end{align}
where the symmetries carry over from real to Fourier space. Reversely, the  generalized structure factor can be obtained evaluating the Fourier coefficients of $G(|\vec{r}-\vec{r\,}'|,z,z')$. For quasi-confinement, the double sum over the mode indices collapses, drastically simplifying the situation. 

\subsection{Integral equation theory}
The static structure factor can be obtained approximately by first introducing the direct correlation function via a suitably adapted Ornstein-Zernike (OZ) relation. Then a closure relation connects the direct correlation function back to the pair-distribution function. Here, we rely on the Percus-Yevick closure, which is particularly simple and also rather accurate for hard spheres~\cite{hansen2013theory,Henderson:Fundamentals_of_inhomogeneous_fluids}.

The Ornstein-Zernike equation adapted for a quasi-confined liquid reads
\begin{align}\label{eq:Ornstein_Zernike}
 h(|\vec{r}-\vec{r\,}'|,z,z') = &c(|\vec{r}-\vec{r\,}'|,z,z') 
  +n \int\mathrm{d}^2r''\int_{-L/2}^{L/2}\mathrm{d}z''
  c(|\vec{r}-\vec{r\,}''|,z,z'') 
  \times h(|\vec{r\,}''-\vec{r\,}'|,z'',z'),
\end{align}
and relates the total correlation function $h(|\vec{r}-\vec{r\,}'|,z,z')=g(|\vec{r}-\vec{r\,}'|,z,z')-1$ to the direct correlation function
$c(|\vec{r}-\vec{r\,}'|,z,z')$. Equation \eqref{eq:Ornstein_Zernike} is a special case of the more general Ornstein-Zernike relation for layered fluids.

By the convolution theorem the OZ relation simplifies in Fourier space
\begin{align}\label{eq:Ornstein_Zernike_Fourier}
 \hat{h}(q,z,z')-\hat{c}(q,z,z')
  &=n\int_{-L/2}^{L/2}\mathrm{d}z''~\hat{c}(q,z,z'')\hat{h}(q,z'',z'),
\end{align}
where we introduced the Hankel transform
\begin{align}\label{eq:Hankel}
    \hat{h}(q,z,z')&=\int\mathrm{d}^2 r e^{i\vec{q}\cdot\vec{r}}h(|\vec{r}|,z,z') \nonumber \\
    &= \int_0^\infty\mathrm{d}r {\mathrm J}_0(qr)r h(r,z,z'),
\end{align}
where ${\mathrm J}_0(\cdot)$ is the Bessel function of the first kind of order 0. Similarly, $\hat{c}(q,z,z')$ is given in the same way.

Here, we supplement the OZ equation by the standard Percus-Yevick closure relation in real space
\begin{align}\label{eq:PY}
 c=f\left(1+h-c\right),
\end{align}
with $f=\exp(-\beta U)-1$ denoting the Mayer
function~\cite{hansen2013theory}. For hard-core interactions the Mayer function evaluates to $-1$ for overlap and $0$ otherwise. Hence, equations~\eqref{eq:Ornstein_Zernike} to \eqref{eq:PY} form a closed set of equations, which can be solved numerically.

The connection of the static structure
factor, equation~\eqref{eq:SF_sim} 
to the total correlation function, equation~\eqref{eq:total_correlation} reads upon decomposition into Fourier modes
\begin{equation}
 S_{\mu}(q)=1+\frac{n_0}{L^2} h_{\mu}(q),
\end{equation}
where only the diagonal component 
\begin{align}
\hat{h}_{\mu}(q) \equiv \hat{h}_{\mu\mu}(q) = \int_{-L/2}^{L/2} \mathrm{d} z &\int_{-L/2}^{L/2} \mathrm{d}z' \hat{h}(q,z,z') 
    \times\exp[ -i Q_\mu (z-z')]
\end{align}
enters.

We solve equations~\eqref{eq:Ornstein_Zernike_Fourier} to \eqref{eq:PY} iteratively. The Hankel transform and its corresponding inverse can be calculated by introducing logarithmic grids. Then equation~\eqref{eq:Hankel} becomes a convolution integral, which can be computed efficiently using a fast Fourier transform~\cite{talman1978fourier,hamilton2000powerspectrum}. We calculate the integrals for distances $r/\sigma$ and wave numbers $q \sigma$ lying within the range $[10^{-4},10^4]$ using 2048 grid points. To avoid an amplification of numerical errors during the iteration, cutoffs $r_{\text{min}}/\sigma=q_{\text{min}}\sigma=2.5\times10^{-3}$ and $r_{\text{max}}/\sigma=q_{\text{max}}\sigma=2.5\times10^{3}$ are introduced, where $h$ and $c$ are put to zero outside of the intervals $[r_{\text{min}}/\sigma,r_{\text{max}}/\sigma]$ and $[q_{\text{min}}\sigma,q_{\text{max}}\sigma]$. The remaining $z$ integral in equation~\eqref{eq:Ornstein_Zernike_Fourier} is calculated using a simple trapezoidal rule on a uniformly discretized grid with spacing $0.01\sigma$. The quasi-confinement is realized by using periodic boundary conditions. 

\subsection{Simulation}

We perform event-driven simulations of hard spheres in three dimensions~\cite{rapaport2004art}, in the periodic geometry illustrated in figure~\ref{fig:torus}. 
We have simulated systems with confinement lengths $L=2.1\sigma$ up to $L=15\sigma$. 
The size of the periodic box in the other two lateral dimensions is approximately $L_{\text{box}}=80\sigma$, depending on the packing fraction and confinement length. This length was chosen to give suitable $q$ resolution in the structure factor calculations. 
We consider densities of $n=$ 0.35, 0.7 and 0.85, corresponding to packing fractions $\varphi := N \pi \sigma^3 /6 A L $ of  $\varphi=0.183,~0.37$ and 0.445. Dependent on $\varphi$, $L_{\text{box}}$ and $L$, the number of particles in each simulation ranges from 7,200-64,000. Due to the hard-sphere interaction, the thermal energy $k_BT$ does not affect the static properties, and enters only via the timescale of the simulation $t_0=\sqrt{m\sigma/k_BT}$, where $m$ is the particle mass.
The simulations are initialized with the particles in a periodic arrangement and velocities assigned randomly, but the system is allowed to equilibrate before any measurements are made. We checked that the results do not change if a longer equilibration period is used. The dynamics at the packing fractions under consideration are not glassy, and the correlations decay on the order of $10t_0$. To ensure adequate statistics, 100 independent runs of each simulation are performed. In each of these repetitions, the simulations are run for a time $10^3t_0$ -- $6\times 10^3t_0$ after the equilibration period, and the configuration is sampled every $50t_0$. These configurations are used to calculate time-averaged static properties of the fluid.

\section{Results and Discussion}\label{sec:results}
\begin{figure}
\includegraphics{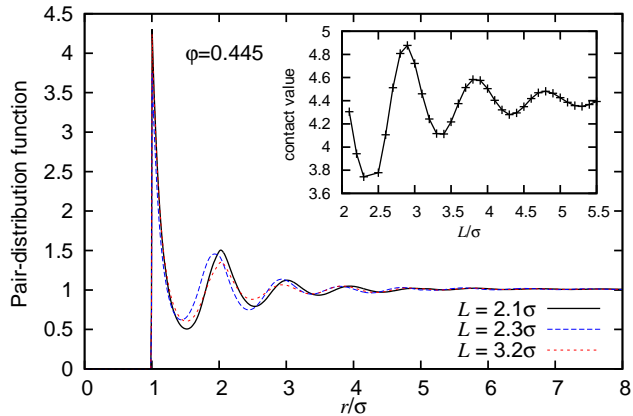}
\caption{\label{fig:RDF} Real-space correlations for three values of the confinement. Pair-distribution function perpendicular to the confinement direction calculated from the simulations at packing fraction $\varphi=0.445$. The insert shows the value of this pair-distribution function at the contact point ($r=\sigma$), as a function of the confinement length $L$. }
\end{figure}

We first consider the real-space correlations, to ensure that our simulations have not crystallized at the packing fractions under consideration. The structure of the fluid can be measured in real space with the pair-distribution function, $g(|\vec{r}-\vec{r\,}'|,z, z')$ given by equation~\eqref{eq:pair_correlation}. 
Due to the symmetry of our system, we choose to sum only over particles which have approximately the same $z$ coordinate, $z=z'$. This correlation is also independent of the direction of $\vec{r}-\vec{r'}$, so we can average over all angles to give us a 2D radial distribution function, measuring the pair correlations perpendicular to the confinement direction. The pair-distribution function shows that the correlations between particles in the simulation die out at long distances, seen in figure~\ref{fig:RDF}, implying that the fluid has not crystallized. Already in this correlation measure we see non-monotonic effects of confinement, demonstrated by the contact value, plotted in the insert of figure~\ref{fig:RDF}. Of the three values of confinement length considered in the main plot, the contact value is  lowest for the middle degree of confinement.

\begin{figure*}
\includegraphics{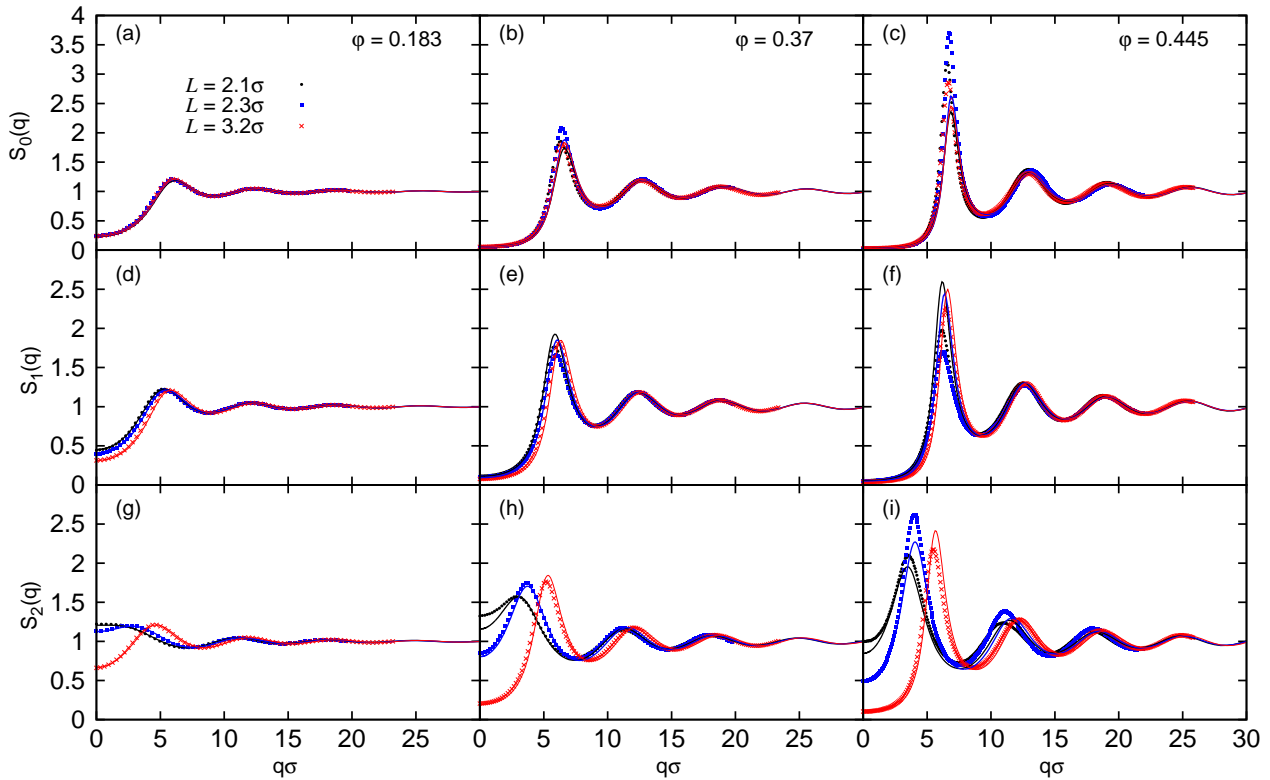}
\caption{\label{fig:SF} Static structure factors $S_\mu(q)$ of the quasi-confined hard-sphere fluid. The first three elements are plotted from the top to bottom panels. The packing fraction $\varphi$ is increased from the left to right. In all cases the simulation results 
are plotted as  points, and the theory as lines of the same color. The most confined liquid, $L=2.1\sigma$, is plotted with black circles, $L=2.3\sigma$  with blue squares and $L=3.2\sigma$ with red crosses.}
\end{figure*}

The main measure we use to characterize the structure of the fluid is the static structure factor, equation~\eqref{eq:SF_sim}. It is a measure of the correlations in the system in Fourier space, and is a natural choice in the study of fluids because it is measurable directly experimentally by scattering experiments. It is also the starting point for dynamic theories of liquids and glasses, such as the mode-coupling theory~\cite{gotze2008complex}. The lowest-mode structure factor  $S_0(q)$ depends only on the lateral coordinates and correspondingly we refer to it as the in-plane structure factor. In contrast, the higher modes $S_\mu(q), \mu > 0$ are sensitive to the arrangement along the confinement direction and reveal valuable information on the competition between local packing and confinement.

\begin{figure}
\includegraphics{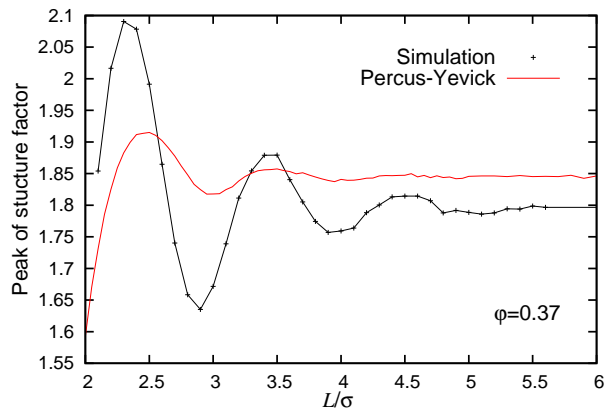}
\caption{\label{fig:SFpeak} Non-monotonic behavior of the peak of the in-plane static structure factor  as a function of confinement length $L$.  The Percus-Yevick  results are plotted with a red line. The simulation results are plotted as black crosses, connected with a line as a guide for the eye. The packing fraction used is $\varphi=0.37$.
}
\end{figure}

At low packing fraction, we find that  the in-plane structure factor  $S_0(q)$ is virtually independent of the confinement length $L$, seen in figure~\ref{fig:SF}(a). 
The effect of confinement can only be seen in higher modes, demonstrated in figure~\ref{fig:SF}(d) and (g). In each of these cases the Percus-Yevick closure is a very close match to the simulations. At higher packing fractions we begin to observe quantitative differences between the structure factor calculated with the PY closure and  simulations, seen in the right two columns of figure~\ref{fig:SF}. Yet, the curve shapes are similar and the wave vector corresponding to each peak matches. Nevertheless, the height of the first peak is quite different, particularly at the highest packing fraction considered. However, it is at these higher packing fractions ($\varphi=0.37$ and 0.445) where we also start to see interesting effects due to the confinement. In particular, the peak of the in-plane structure factor shows a non-monotonic dependence on the confinement length, seen in figure~\ref{fig:SFpeak}. The curves for the simulation and PY closure  are qualitatively similar, with oscillations occurring at almost the same values of $L$, but the oscillations are significantly less pronounced in the theoretical  results. At large values of $L$ the structure factor peak appears to saturate at  a constant value for both simulation and theory, as expected for approaching the bulk. 
Consistent with the literature, the PY approximation overestimates the height of the first peak of the static structure factor in the bulk~\cite{hansen2013theory}. The height of the main peak of the structure factor is a measure of the near-ordering of the fluid. The oscillatory behavior of this peak suggests that the degree of ordering in the fluid changes from commensurate to incommensurate packing. This interpretation is corroborated by the fact that the period of the oscillates coincides with the  hard-sphere diameter. 
In the simulation, the maximum appears at $L=2.3\sigma$. This is the same behavior as liquids confined between hard walls~\cite{mandal2014multiple}.

\begin{figure}
\includegraphics{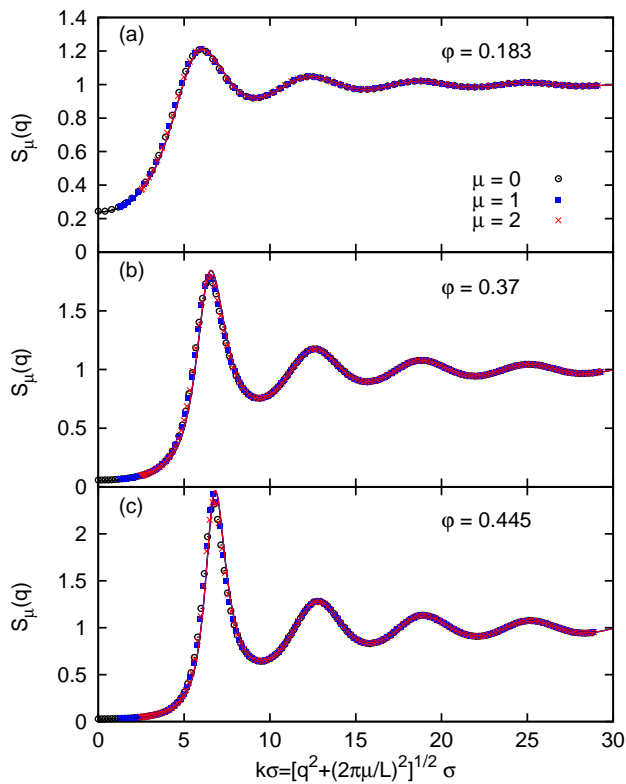} 
\caption{\label{fig:SFbulk} Static structure factors at large confinement length, $L=5\sigma$, 
plotted as a function of the length of the 3D wavevector $k= \sqrt{q^2 + Q_\mu^2}$.
The first three modes of the structure factor are included, plotted as black circles, blue squares and red crosses respectively. The simulation results are included as points and the theory results as lines. The packing fraction is increased from top to bottom, with $\varphi=0.183$ in (a), $\varphi=0.37$ in (b) and $\varphi=0.445$ in (c). Note that in each case all modes of the structure factor overlap.}
\end{figure}

The constant value of the peak of the in-plane static structure factor at large $L$ implies that the system is approaching the bulk limit. Then we expect that  3D rotational invariance is restored such that only the magnitude of the 3D wave vector $\vec{k} = (\vec{q}, Q_\mu)$ determines the structure factor
\begin{align}
S_\mu(q) \to S(k) \qquad \text{with } k =\sqrt{q^2 + Q_\mu^2},
\end{align}
as $L\to \infty$. 
This convergence is confirmed in figure~\ref{fig:SFbulk}
for $L=5\sigma$ for the lowest three modes of the structure factors.  
We can also observe from this plot that while the differences between simulation and the PY closure are smaller than in the quasi-confined fluids with higher degrees of confinement, they still grow with increasing packing fraction, consistent with  bulk behavior~\cite{hansen2013theory}.

\begin{figure}
\includegraphics{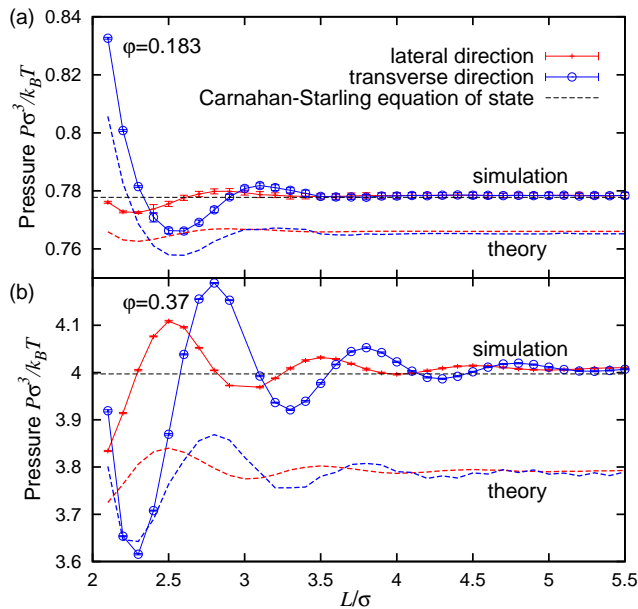}
\caption{\label{fig:Pressure} Transverse ($P_{zz}$) and lateral ($P_{xx}$) components of the pressure tensor as a function of confinement length $L$. The simulation results are included as points connected by a solid line, and the corresponding theory results from the PY closure are plotted as a dashed line of the same color. These values are calculated at packing fraction $\varphi=0.183$ in (a) and $\varphi=0.37$ in (b). The bulk pressure calculated from the Carnahan-Starling equation of state is included as a dashed black line for comparison.}
\end{figure}

Given the non-monotonic behaviour of the pair-distribution function and static structure factor, we would also expect to see oscillations in the pressure of the fluid as a function of confinement length. For a hard-sphere fluid without rotational symmetry, the pressure is measured by the diagonal components of the pressure tensor in the transverse and lateral directions~\cite{de2006detailed}. Its form can be derived starting from thermodynamics, and was presented already by Irving and Kirkwood~\cite{irving1950statistical}. In terms of inter-particle forces, this expression is given by
\begin{align}
P_{zz}& = n k_BT- 
\frac{n^2}{2}   \int_{A}\int_{-L/2}^{L/2}   \frac{z^2}{\sqrt{r^2+z^2}} u'(\sqrt{r^2+z^2}) g(r,z) \textrm{d}z \textrm{d}^2 r  
\end{align}
for the transverse component, in our quasi-confined geometry. Here, $u$ is the pair potential and $g(r,z)$ is the pair-distribution function, which  depends only on the in-plane and transverse distances $r$ and $z$. Following the approach outlined in Hansen and McDonald~\cite{hansen2013theory}, this can be written in a form suitable for calculation in a hard-sphere fluid as 
\begin{align}\label{eq:Pzz}
P_{zz}  &=  n k_BT+ k_BT \pi   n^2 \int_{-\sigma}^{\sigma}      g(\sqrt{\sigma^2-z^2},z)z^2 \textrm{d}z .
\end{align}
Here the pair-distribution function has to evaluated at contact, while the $z$-integral corresponds to averaging over the surface of the sphere at contact. Similarly, for the lateral pressure we find 
\begin{align}\label{eq:Pxx}
P_{xx}  &=  n k_BT+ k_BT \frac{\pi   n^2}{2} \int_{-\sigma}^{\sigma} g(\sqrt{\sigma^2-z^2},z)    (\sigma^2-z^2)  \textrm{d}z.
\end{align}
Both pressures   can be handled numerically  using Simpson's rule with the pair-distribution function obtained within PY.
 The transverse component of the pressure is rather  sensitive to the $z$-discretization of the $g$ calculation, so we use a finer grid spacing of $0.001\sigma$ for $\varphi=0.183$ and $0.002\sigma$ for $\varphi=0.37$.

In our simulations, the components of the pressure tensor are calculated from the collisions of the spheres~\cite{henderson1984interface} as 
\begin{align}
P_{zz}& = n k_BT-\frac{m}{A L} \left\langle\frac{1}{t_m}\sum_{c} \frac{z_{ij}^2}{\sigma^2}(\vec{r}_{ij}\cdot\vec{v}_{ij}+z_{ij}v_{zij}) \right\rangle  
\label{eq:PressureSim}
\end{align}
for the transverse component. The sum runs over all collisions $c$ occurring in a time interval $t_m$. The indices $i$ and $j$ refer to the particles involved in the collision, with the difference in their $x$--$y$ position at the instant of collision denoted by $\vec{r}_{ij}$, the difference in the $z$-component of the positions by $z_{ij}$, the difference in their $x$--$y$ velocity by $\vec{v}_{ij}$, and the difference in the $z$ component of their velocity by $v_{zij}$.
The  lateral component $P_{xx}$ is calculated similarly. Both components oscillate as a function of $L$, plotted in figure~\ref{fig:Pressure}. The confinement has a larger effect on the transverse component of the pressure, evident by the larger oscillations, seen for both the simulation and PY result. At large values of $L$ both the lateral and transverse components approach the same constant value, consistent with the fluid approaching the bulk state. The simulation value is approximately $0.3\%$ above the one calculated with the Carnahan-Starling equation of state~\cite{carnahan1969equation} for a bulk hard-sphere fluid, which is within its expected accuracy~\cite{hansen2013theory}. The PY value, for both packing fractions considered, does not quantitatively match the simulation result. However, qualitatively the curves look similar; the oscillations in each component occur at the same $L$ values, for both packing fractions. 

Notably, the oscillations in the two components of the pressure tensor are out of phase. As the lateral pressure spikes, it is partially compensated for by a dip in the transverse pressure. The location of the peaks is dependent on the packing fraction, but the period of the oscillations is in both cases equal to the particle diameter. This matches the oscillations seen in the static structure factor, and indicates a dependence on commensurate or incommensurate packing.

\section{Conclusions}\label{sec:conclusions}
We have investigated the effect of confining a fluid by applying extremely small periodic boundary conditions in one direction. Through the use of event-driven simulations and integral equation theory within the Percus-Yevick closure we have characterized the structure of the fluid through the generalized static structure factor. We see that the correlations in the structure show a non-monotonic dependence on the confining length scale, in a very similar manner to a liquid confined between two plates, even though the density profile of our fluid is constant. This indicates that the correlations in confined liquids may not be intrinsically related to the oscillating density profile they exhibit. 

Additionally, we have seen that the non-monotonic behavior of the fluid can be observed in the pair-distribution function, and the components of the pressure tensor. As the confinement length is increased, the bulk isotropic fluid is approached, indicated by both components of the pressure tensor approaching the same value, and all elements of the structure factor overlapping.

While the Percus-Yevick approximation gives qualitatively the same results as the simulations, there are significant discrepancies between the structure factors calculated. One may seek to improve the solution empirically with the simulation data, in a manner similar to the Verlet-Weis correction~\cite{verlet1972equilibrium} used in the bulk. However, this method relies on the Carnahan-Starling equation of state, for which we have no equivalent in the quasi-confined fluid. As such, deriving the viral expansion of the equation of state in this system could be of value. Based on our calculations of the pressure, it will clearly be a complicated function of confinement length.

The present findings motivate further investigations into quasi-confined fluids. In particular, the dynamics close to the glass transition are of interest, as we expect this system provides a means to investigate the interplay of relevant length scales and incommensurability effects on the glass transition. Such an investigation is in a similar vein to investigations into the glass transition that considered the effects of curvature on local particle packing, by confining liquids to the surface of a sphere~\cite{vest2015mode,vest2014dynamics,vest2018glassy} or hypersphere~\cite{turci2017glass}. There, non-monotonic behavior of the correlations in the structure was observed for particles confined to the surface of a hypersphere with a circumference between 10 and 20 particle diameter lengths. A further extension to our work could include confining fluids to hyperspheres of smaller size. This would have the advantage of even higher symmetry than the periodic quasi-confinement, and could reveal an interplay between the effects of quasi-confinement and curvature.  In order to study fluids at a higher packing fraction, in the vicinity of the glass transition, it would be necessary to use polydisperse hard spheres in the simulations. We expect that mode-coupling theory dynamics of the monodisperse hard spheres could be compared to simulations of slightly polydisperse systems~\cite{van1994glass}, as is done successfully in the bulk~\cite{weysser2010structural}. The results presented here have demonstrated that the non-monotonic behavior of pair correlations in confined liquids is not simply a reflection of the layering. This extension would seek to disentangle the non-monotonic behavior of the glass transition from the inhomogeneous density profile of confined fluids. 

\begin{acknowledgments}
We thank Martin Oettel for introducing us to the numerics of integral equation theory. 
CFP is supported by the Austrian Science Fund (FWF): M 2471. TF and LS are supported by the
FWF: I 2887. The computational results presented have been achieved in part using the HPC infrastructure LEO of the University of Innsbruck. 
\end{acknowledgments}

\bibliography{torus}

\end{document}